\documentstyle[11pt,amsfonts]{article}
\renewcommand{\theequation}{\arabic{equation}}
\newcommand{\be}{\begin{equation}}
\newcommand{\ee}{\end{equation}}
\newcommand{\bea}{\begin{array}}
\newcommand{\ea}{\end{array}}
\newcommand{\beqa}{\begin{eqnarray}}
\newcommand{\eeqa}{\end{eqnarray}}
\newcommand{\bean}{\begin{eqnarray*}}
\newcommand{\eean}{\end{eqnarray*}}

\def\up#1{\leavevmode \raise.16ex\hbox{#1}}

\newcommand{\gapproxeq}{\lower
 .7ex\hbox{$\;\stackrel{\textstyle >}{\sim}\;$}}
\newcommand{\lapproxeq}{\lower .7ex\hbox{$\;\stackrel
{\textstyle <}{\sim}\;$}}

\renewcommand{\theequation}{\thesection.\arabic{equation}}

\newcounter{appendice}
\newcommand{\appendice}
{
\setcounter{equation}{0}
\renewcommand{\theequation}{\Alph{appendice}.\arabic{equation}}
\addtocounter{appendice}{1}
{\bf Appendix \Alph{appendice}}
}

\def\thebibliography#1{{\bf REFERENCES\markboth
 {REFERENCES}{REFERENCES}}\list
 {[\arabic{enumi}]}{\settowidth\labelwidth{[#1]}\leftmargin\labelwidth
 \advance\leftmargin\labelsep
 \usecounter{enumi}}
 \def\newblock{\hskip .11em plus .33em minus -.07em}
 \sloppy
 \sfcode`\.=1000\relax}

\def\BI{{\rm 1\!l}}

\begin{document}


\centerline{ \LARGE Can classical wormholes stabilize the
  brane-anti-brane system? } 

\vskip 2cm

\centerline{ {\sc    A. Pinzul$^{a}$ and A. Stern$^{b}$ }  }

\vskip 1cm
\begin{center}
{\it a)  Department of Physics, Syracuse University,\\ Syracuse, 
New York 13244-1130,  USA\\}
{\it b) Department of Physics, University of Alabama,\\
Tuscaloosa, Alabama 35487, USA}

\end{center}

\vskip 2cm

\vspace*{5mm}

\normalsize
\centerline{\bf ABSTRACT} 
We investigate  the static
solutions  of Callan and Maldecena and Gibbons
 to lowest order Dirac-Born-Infeld theory.  Among them are  charged 
 wormhole solutions
 connecting  branes to  anti-branes.  It is seen that
 there are no such solutions when the separation
  between the brane and anti-brane is  smaller than some
  minimum value.  The minimum
  distance coincides with the energy minimum, and    depends monotonically on the charge.   Making the charge 
  sufficiently large, such that the minimum separation is much bigger than $
  \sqrt{\alpha '}$, may suppress known quantum processes leading to  decay of the
  brane-anti-brane system.  For this to be possible the zeroth order wormhole
  solutions should  be reasonable approximations of solutions in the
  full $D-$brane theory.  With this in mind we address the question of  whether
  the zeroth order solutions are stable under inclusion of higher order
  corrections to the Dirac-Born-Infeld action.

\vspace*{5mm}

\newpage
\scrollmode

\section{Introduction}

The Born-Infeld  nonlinear description of electrodynamics\cite{bi}
and its subsequent
generalization to membranes \'a la Dirac\cite{Dirac} is of current
interest due to its role as an effective theory for
D$p-$branes.  The associated Dirac-Born-Infeld (DBI) action appears at  lowest order  in the derivative expansion
for the effective  D$p$-brane action.\cite{Fradkin:1985},\cite{Johnson} 
 The original  Born-Infeld theory has a charged static solution, which
 was generalized by Callan and Maldecena\cite{Callan:1997kz}
 and Gibbons\cite{Gib} to   families of static solutions
 on the brane.  The families are associated with orbits of the
 $SO(1,1)$ group.   The Lagrangian is invariant under $SO(1,1)$ and
 can be used to label the orbits.  One such orbit contains the
 solution of  Born
 and Infeld.  Another is a BPS solution representing a fundamental string attached to the
brane.  Finally, there is  a third family of  solutions
corresponding to wormholes which connect the brane to an anti-brane a
finite distance away.  Here we show
that there are no charged wormhole solutions having a separation
  between the $p-$brane and anti$-p-$brane  smaller than some
  minimum value.  The  minimum separation
 distance goes like $|Q|^{\frac 1{p-1}}$, $Q$ being the $U(1)$
 charge.  At minimum separation, the self-energy
 is also a minimum, where there appears  a cusp singularity in the
 plot of the
 energy versus separation distance.  
 In the quantum  analysis of the  brane-anti-brane system an instability is known to occur at distance scales of order $ \sqrt{\alpha '}$ due to
  excitation of tachyonic modes.\cite{Banks:1995ch}
  Then for sufficiently large charge, i.e.
 \be |Q|^{\frac 1{p-1}}>> \sqrt{\alpha '}\;,
  \ee such  quantum processes may be suppressed, and it is
  possible that
  charged wormholes can stabilize the  brane-anti-brane
 system. 

For the above scenario to be correct, however, classical stability of
the wormhole  solutions should be checked.  This means  $a)$ enlarging to
time dependent solutions, and investigating whether solutions are
stable with respect to perturbations about the
static solution.  It also means $b)$ checking whether the solutions to  the zeroth order
effective theory are a reasonable approximation of solutions to the
full effective  D$p-$brane action.   Here we shall only consider $b)$.
One signal that  solutions may be unstable in the sence $b)$ is the presence of singularities in the
 field strength, where  the  derivative expansion cannot be trusted.  Such a singularity is present for the original
Born-Infeld solution, in that case at a single point, and for the
entire orbit of solutions connected to the Born-Infeld solution.  Despite the
singularity, these
solutions are associated with a
finite self-energy.  For the BPS case the
singularity occurs an infinite distance away from the brane, and
appears harmless.   The  wormhole-type solutions  were originally
constructed by  joining together  two local solutions, obtained in the static gauge,
 at the minimum circumference of the wormhole.\cite{Callan:1997kz},\cite{Gib}
A singularity in the field strength occurs along the throat - precisely  at the minimum
circumference.   The singularity in the field strength is a
coordinate singularity, which can be removed by going to another gauge.
Nevertheless, it is a signal that higher order corrections may not be negligible.

 To check stability in the sence $b)$, we will rely on recent
 computations of the derivative corrections to Born-Infeld theory.
 The first order corrections to the action  were obtained separately  by Wyllard \cite{Wyl}  and
 Das, Mukhi and Suryanarayana\cite{das}. [Higher order corrections
 seem currently out of reach.]  Using their results we
 carried  out a stability check previously for the case of the original Born-Infeld
solution.\cite{Karatheodoris:2002bb}  There we argued that the original
 Born-Infeld solution   is  unstable under inclusion of 
these first order corrections.     More specifically, we numerically obtained
 corrections to the zeroth order Born-Infeld solution, but found that
they give an infinitely large correction to the Lagrangian.
  We give a simpler proof of the  result here.  Because the Lagrangian is $SO(1,1)$
 invariant the result applies to the entire orbit of solutions
 connected to the Born-Infeld solution.   Concerning the BPS solution,
 it is known that the such a solution is stable {\bf to all
  orders} in the derivative expansion.\cite{Thorlacius:1997zd}  We
verify that this is consistent with the first order results of \cite{Wyl} and
 \cite{das}.   We find that the stability analysis for the wormhole solutions
 leads to the same results obtained for the Born-Infeld solution.
 Namely,  corrections to the zeroth order  solution lead to an
 infinitely large correction of the Lagrangian.  In this case, we need
 to rely on numerical computation for the result.

In section 2 we
give analytic expressions for the  three families of solutions along with their
self-energies.  Here we show that the charged  wormhole solutions
have a minimum length.
The question of  stability of the zeroth order  solutions is addressed in section 3.   In  appendix A we write down the wormhole solution in another gauge, where
 the field strength is singularity-free.  In fact, it is a constant in
 that gauge.   In appendix B  we  use the results of \cite{Wyl}  and
  \cite{das} to obtain the first order corrections to
 the field equations for the BPS case, and show that the answer  agrees with   \cite{Thorlacius:1997zd}.  

\section{Zeroth Order Solutions}
\setcounter{equation}{0}
\subsection{Dirac-Born-Infeld theory}

We consider the $p$-dimensional brane embedded in a ten dimensional
space-time with flat metric $[\eta_{AB}]=$diag$(-1,1,...,1)$, $A,B,..=0,1,...,9$.  We denote the
brane coordinates by  $X^A$.  They are functions of  $p+1$ parameters
$\xi^\mu$, $\mu,\nu,...=0,1,...,p$.   Additional degrees
of freedom on the brane  are $U(1)$ potentials ${\cal A}_\mu(\xi)$.  
The DBI action is written in terms of the $(p+1)\times(p+1)$ matrix
\be  h_{\mu\nu} =\eta_{AB}\;\partial_\mu
X^A(\xi)\;\partial_\nu X^B(\xi)+2\pi\alpha'\; F_{\mu\nu}(\xi)  \;, \label{giffh}\ee 
where $\partial_\mu =\frac{\partial}{\partial \xi^\mu}$.  The first
term is the induced metric on the brane, while  $
F_{\mu\nu}=\partial_\mu{\cal A}_\nu-\partial_\nu{\cal A}_\mu$ is the $U(1)$
field strength.  We assume the two-form contribution is absent
$B_{AB}=0$.    The DBI action  is\cite{bi},\cite{Dirac}
\be {\cal S}^{(0)}_{DBI} =  \frac{T_p}{ g_s} \int d^{p+1}\xi\; {\cal
  L}^{(0)}_{DBI} \;,\qquad {\cal L}^{(0)}_{DBI} = 1-\sqrt{-\det [
  h_{\mu\nu}]}\;, \label{Dpact} \ee where
$T_p$ is the tension, which is expressable in terms of $\alpha'$
according to \be T_p ={(4\pi^2 \alpha ')^{-(p+1)/2} }\;,\label{Tp}\ee and
$g_s$ is the string coupling.
$ {\cal S}^{(0)}_{DBI}$ is invariant under diffeomorphisms on the brane $\xi^\mu \rightarrow
\xi'^\mu(\xi)\;,$  $U(1)$ gauge transformations ${\cal A}_\mu(\xi) \rightarrow
{\cal A}_\mu(\xi)  +\partial_\mu\Lambda(\xi) $, as well as 
ten-dimensional Poincar\'e transformations.  Variations in $X^A(\xi)$
and ${\cal A}_\mu(\xi)$ lead to the equations of motion
\beqa  \partial_\mu\biggl\{ \sqrt{-\det 
  h}\;(h^{\mu\nu}+ h^{\nu\mu}) \eta_{AB} \partial_\nu X^B\biggr\}&=& 0 \cr & &\cr
\partial_\mu\biggl\{ \sqrt{-\det 
  h}\;( h^{\mu\nu}- h^{\nu\mu})\biggr\}&=& 0 \;,
\label{xaeom}\eeqa respectively,
where $h^{\mu\nu}h_{\nu\rho} = \delta ^\mu_\rho$.

The known families of spherically
symmetric static  solutions\cite{Callan:1997kz},\cite{Gib} can be
classified in terms of $SO(1,1)$ orbits (we do this below), and they
  describe different topologies  embedded in the flat
  ten-dimensional background.  For one family of solutions,  containing the original
  Born-Infeld solution, 
a time slice is ${\mathbb{R}}^p$ minus a point.     We
could therefore describe it with the introduction of a delta function source to the right-hand-side
of (\ref{xaeom}).  Another family
  corresponds to a
brane and anti-brane connected by a wormhole.   In that case one can patch together local solutions to
(\ref{xaeom}).  The families of solutions are written in terms of two
integration constants $Q$ and $C$, $Q$ being the electric charge.  
Locally, all solutions can be expressed in the static gauge, where one identifies $\xi^\mu$ with the first
$p+1$ brane coordinates $X^\mu$, $\mu=0,1,...,p$.  The remaining  $X^\alpha$,    $
\alpha= p+1,p+2,...,9$,  then denote normal coordinates, and 
(\ref{giffh}) becomes
\be  h_{\mu\nu} =\eta_{\mu\nu} +
\;\partial_\mu X^\alpha\partial_\nu X^\alpha +2\pi\alpha'\; F_{\mu\nu}\;, \label{hisg}\ee
The static solutions of \cite{Callan:1997kz},\cite{Gib}  are for  a radial electric field
with a single transverse mode  excited.   Choose the    nonvanishing degrees of freedom to be ${\cal
  A}_0(r)$ and $X_{p+1}(r)$, where $r$ is the radial coordinate on the
brane.   Since the metric is diagonal, the resulting matrix $h$ is diagonal except for the $2\times 2$
submatrix with corresponding  indices $\mu$ and $\nu$ equal to $0$ and $r$.  That  $2\times 2$
submatrix and its inverse are given by
\be  \pmatrix{-1  &  -f(r) \cr f(r)  &1+g(r)^2  \cr} \qquad{\rm and}\qquad
\frac1{1+g(r)^2 -f(r)^2} \pmatrix{-1-g(r)^2  &  -f(r) \cr f(r)  &1
  \cr}\;, \ee  respectively, where $f(r)=2\pi\alpha'\partial_r{\cal
  A}_0(r)$ and $g(r) =\partial_rX_{p+1}(r)$.   Substituting into
the equations of motion (\ref{xaeom}) gives
\be \partial_r\biggl\{\frac {r^{p-1}f(r)}{\sqrt{1+g(r)^2 -f(r)^2}}\biggr\} =
\partial_r\biggl\{\frac {r^{p-1}g(r)}{\sqrt{1+g(r)^2 -f(r)^2}}\biggr\} =0 \;\ee
The solutions for $f(r)$ and $g(r)$ are
\be  \frac{f(r)}Q  = \frac{g(r)}C  = \frac{1}{\sqrt{Q^2 -C^2
    +r^{2p-2}}} \label{sln0} \ee 
The integration constants $Q$ and $C$ have units of [length$]^{p-1}$.
 
For the configurations (\ref{sln0}) the Lagrangian and equations of motion are invariant under the
$SO(1,1)$ transformation 
\be \pmatrix{ f(r) \cr g(r)}\rightarrow \pmatrix{\cosh\nu &\sinh \nu\cr
\sinh\nu &\cosh \nu\cr}\pmatrix{ f(r) \cr g(r)}\label{so11}\ee
The integration constants  transform in the same
way  \be \pmatrix{ Q \cr C}\rightarrow \pmatrix{\cosh\nu &\sinh \nu\cr
\sinh\nu &\cosh \nu\cr}\pmatrix{ Q \cr C}\label{so11qc}\ee
There are then three kinds of orbits: $i)\;|Q|>|C|$, $ii)\;|Q|=|C|$ and $iii)\;|Q|<|C|$. $i)$ is connected to the original
Born-Infeld solution,  $ii)$ is the BPS solution and $iii)$ is associated
with wormhole solutions.   The orbits can be classified by their corresponding  value for the spatial
  integral of the Lagrangian density $ {\cal
  L}^{(0)}_{DBI}$  
\be   L^{(0)}_{DBI}(r)=-  \Omega_{p-1}\biggl\{\int^\infty_r dr  
 \frac{ r^{2p-2} } {\sqrt{ r^{2p-2} +Q^2-C^2}}-\int^\infty_0 drr^{p-1} \biggr\}
  \;, \ee  where $ \Omega_{p-1}$ is the volume of a unit
  $p-1$ sphere.  We  removed  a hole around the origin of
  radius $r$ in the integration for the first integral, in order to
  accommodate the different cases, as
  their domains differ.  The second integral is the  vacuum subtraction.  The first integral
  can be expressed in terms of the one for  $X_{p+1}(r)$, yielding
\be   L^{(0)}_{DBI}(r)=\frac {  \Omega_{p-1}}p
\biggl\{ r\sqrt{ r^{2p-2} +Q^2-C^2}\; +\; (C^2-Q^2)\frac{X_{p+1}(r)}C
 \biggr\}
 \label{lfbih} \ee  For cases $i),\; ii)$ and $iii)$ we find that the
 result (after setting $r$ to its appropriate value) is positive, zero and negative, respectively. 

\subsection{Analytic solutions}

The right hand side of (\ref{sln0}) can be integrated to obtain
analytic expressions for the potential $ {\cal A}_{0}(r)$ and the
transverse displacement $
X_{p+1}(r)$.  For this  expand  in powers of  $ \frac
{r^{2p-2}}{Q^2-C^2}$ and integrate term by term.  The result  for the
indefinite integral
 can be expressed in terms of a hypergeometric
function:\be
 \frac {r}{\sqrt{Q^2-C^2}}\;
F\biggl(\frac  12,\frac 1 {2p-2};\frac {2p-1} {2p-2};- \frac
{r^{2p-2}}{Q^2-C^2} \biggr)\;,\label{indef}\ee  for  $ |\frac
{r^{2p-2}}{Q^2-C^2}|<1$.
Now set the limits of integration to be $r$ and $\infty$, with the
assumption that the potentials vanish at the latter.  To evaluate 
(\ref{indef}) for these limits we  analytic
continue  $F(a,b;c;z)$ using\cite{Bateman} 
\beqa F(a,b;c;z)&=&\frac {\Gamma(c)\Gamma(b-a)}
{\Gamma(b)\Gamma(c-a)}(-z)^{-a}\;F(a,1-c+a;1-b+a;\frac1z)\cr 
&+&\frac {\Gamma(c)\Gamma(a-b)}
{\Gamma(a)\Gamma(c-b)}(-z)^{-b}\;F(b,1-c+b;1-a+b;\frac1z)\label{idfhyg}\eeqa
where $|{\rm arg}(-z)|<\pi$.  So (\ref{indef}) can be rewritten as
$$ \frac1{(2-p)r^{p-2}}\;F\biggl(\frac12,
\frac{p-2}{2p-2};\frac{3p-4}{2p-2};-\frac{Q^2-C^2}
{r^{2p-2}} \biggr)\;+\;
\frac{(Q^2-C^2)^{\frac{2-p}{2p-2}}}{\sqrt\pi}\;\Gamma\biggl(\frac{2p-1}{2p-2}\biggr)
\Gamma\biggl(\frac{p-2}{2p-2}\biggr)\;,
$$ since $F(a,0;c;z)=1$.  For $p>2$, only the last term survives when evaluating
at $\infty$, which then get subtracted out after evaluating
between $r$ and $\infty$.  For the transverse displacement 
 $X_{p+1}$ one gets 
\be
X_{p+1}(r) =  \frac{-  C} { (p-2)r^{p-2}}\;F\biggl(\frac12,
\frac{p-2}{2p-2};\frac{3p-4}{2p-2};-\frac{Q^2-C^2}
{r^{2p-2}} \biggr), \qquad{\rm for}\;p>2\label{slfx}
\ee

Next we examine the result for the three  different orbits:

  Case  $i)\;|Q|>|C|$.
The limit $C\rightarrow 0$ where the transverse mode is not excited gives the
original Born-Infeld solution\cite{bi}, while $|Q|> |C|>0$ yields a deformation
of the Born-Infeld solution where a spike protrudes from the brane.  We
plot  below the function $X_{p+1}(r) $ for $p=3$ on a two dimensional
spatial slice:

\bigskip
\input epsf
\def\epsfsize#1#2{0.8#1}
\centerline{\hss{\epsffile{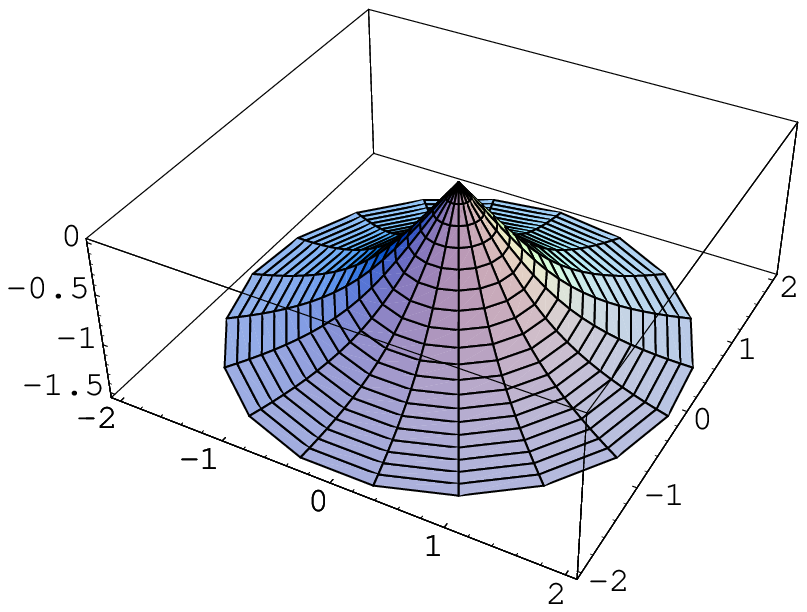}}}

\medskip
\centerline{\qquad\qquad\qquad
${\tt fig. 1}\qquad p=3\qquad |Q|=1\quad |C|=.8$}
\noindent
>From (\ref{slfx}) [and (\ref{idfhyg})] the maximum size of the spike is the
 absolute value of
\be X_{p+1}(0) =\frac{C \;  \Gamma(\frac{3p-4}{2p-2})
\Gamma(\frac{1}{2p-2}) }     {  (2-p)\sqrt{\pi}   }\;(Q^2 -C^2)^{\frac{2-p}{2p-2}}\;\;,
 \qquad{\rm for}\;p>2 \label{xo}\ee   For the integral  of  the
 Lagrangian density [C.f. (\ref{lfbih})] 
one gets \be
 L^{(0)}_{DBI}(0)=\frac{\Omega_{p-1} \;
 \Gamma(\frac{3p-4}{2p-2} )\Gamma(\frac 1 {2p-2})   }{p(p-2)\sqrt{\pi}
 }\;(Q^2 -C^2)^{ \frac p {2p-2} }\; >\;0\;,
 \qquad{\rm for}\;p>2 
\ee

 Case  $ii) \;|Q|=|C|$. One
arrives at the BPS solution  for this case, where (\ref{sln0}) reduce to the Coulumb solutions $f(r)  = g(r)  =
\frac{Q}{r^{p-1}}$.  Since $F(a,b;c;0)=1$, (\ref{slfx}) reduces to \be
 X_{p+1}(r) = \frac{- Q}{ (p-2)\; r^{p-2}}\;, \qquad {\rm
 for}\;p> 2\ee  and
 the  spike
becomes infinitely long, 

\bigskip
\input epsf
\def\epsfsize#1#2{0.8#1}
\centerline{\hss{\epsffile{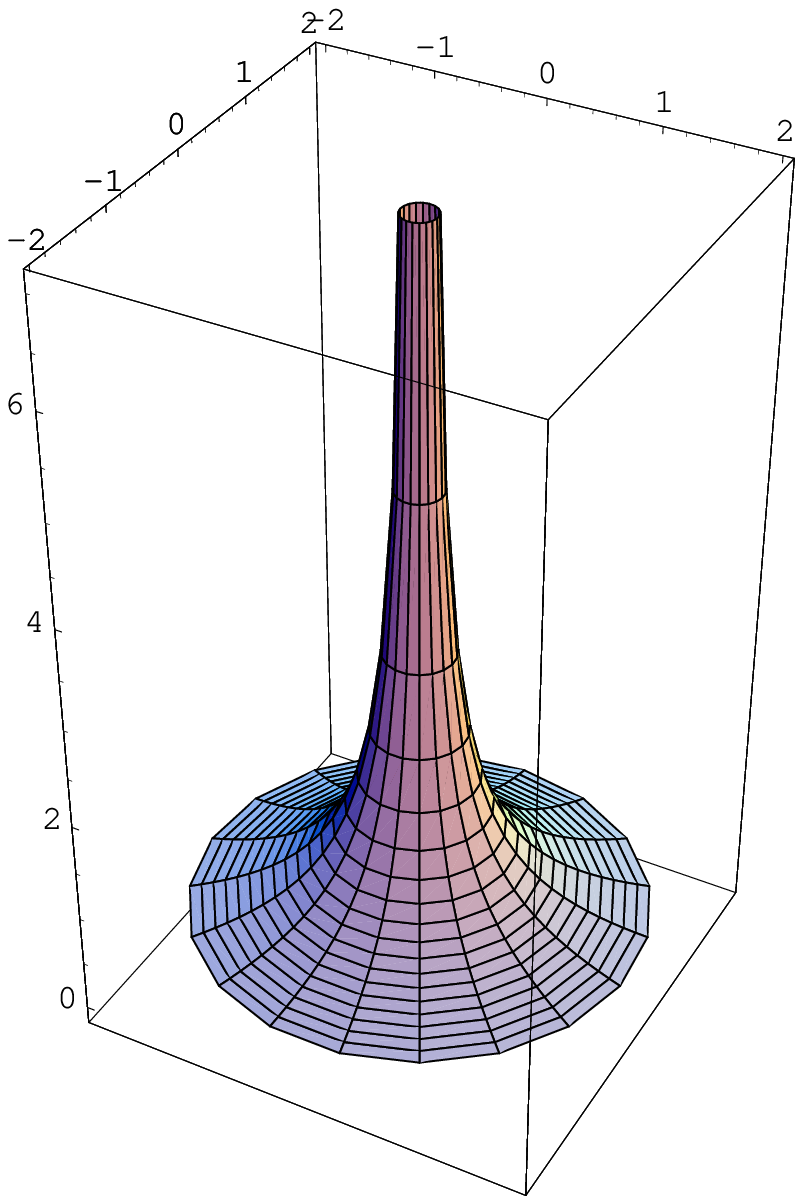}}}

\medskip
\centerline{\qquad\qquad \qquad
${\tt fig. 2}\qquad p=3\qquad |Q|= |C|=1$}
\noindent
representing a fundamental string attached to the
brane.\cite{Callan:1997kz}  In
this case    the integral of  the
 Lagrangian density  
$ L^{(0)}_{DBI}(0)$ goes to zero.
 These solutions are  BPS because they
preserve
half of the supersymmetries of the ground state solution.
Supersymmetries are present when the matrix \be\partial_\mu
X_{p+1}(\xi)\;[\Gamma^\mu,\Gamma^{p+1}]\;+2\pi\alpha'\;F_{\mu\nu}(\xi)\;[\Gamma^\mu,\Gamma^\nu]\label{susycd}\ee
is   degenerate.  $\Gamma^A$  are  $\Gamma$ matrices for the
ten-dimensional background space, $[\Gamma^A,\Gamma^B] = 2\eta^{AB}$.
 To see that this holds when  $|Q|=|C|$, one observes that
(\ref{susycd}) is proportional to $(Q\Gamma^0 +
C\Gamma^{p+1})\Gamma^r$, whose square is
$(C^2-Q^2)\BI$.  (\ref{susycd}) is then nilpotent when $|Q|=|C|$.  

 Case  $iii)\;|Q|<|C|$.  Here one  gets a finite diameter  tube  with a minimum radius $r_0 =
(C^2-Q^2)^{\frac1{2p-2}}$:

\bigskip
\input epsf
\def\epsfsize#1#2{0.8#1}
\centerline{\hss{\epsffile{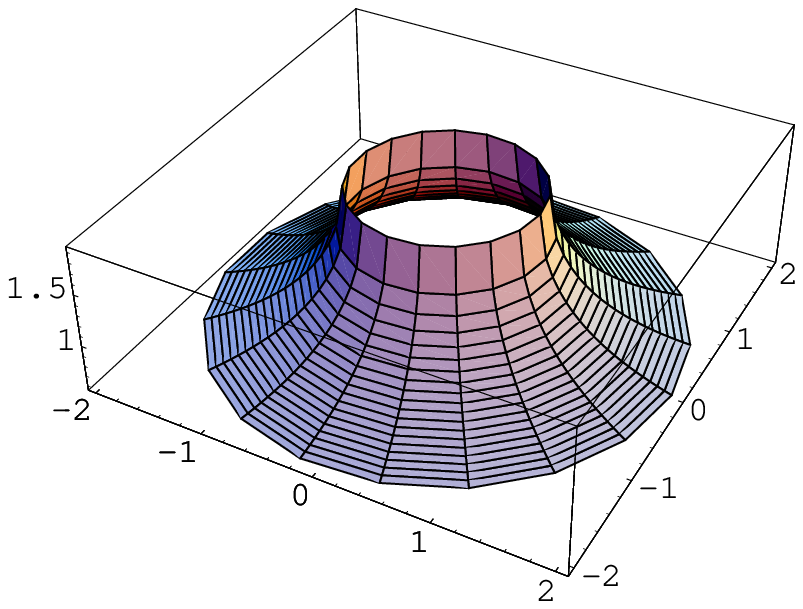}}}

\centerline{\qquad\qquad\qquad
${\tt fig. 3}\qquad p=3\qquad |Q|=1\quad |C|=1.2$}
\noindent
Both
 $  g$ and $f$, and consequently also the electric field, are singular at
 $r=r_0$.  Nevertheless, ${\cal A}_0$ and $ X_{p+1}$ are not.  From
 the latter the
 tube has a finite length.  After
 expressing $C$ in terms of $r_0$ and $Q$, it is  
 \be X_{p+1}(r_0) = \frac
 {\sqrt{\pi(r_0^{2p-2}+Q^2)}}{ (2-p)\;r_0^{p-2}}\;\frac{
   \Gamma(\frac{2p-1}{2p-2} )}{
\Gamma(\frac{p}{2p-2})}\;, \qquad{\rm for}\;p>2 \;,\label{xro}\ee 
 The domain of integration for the integral  of  the
 Lagrangian density [C.f. (\ref{lfbih})] now goes from $r_0$ to
 $\infty$.  One gets \be
 L^{(0)}_{DBI}(r_0)=-\frac{\Omega_{p-1} \;\sqrt{\pi}}{p(p-2)}
\;\frac{
   \Gamma(\frac{2p-1}{2p-2} )}{
\Gamma(\frac{p}{2p-2})}\;r_0^p \;< \;0\;,
 \qquad{\rm for}\;p>2 
\ee

 The static gauge breaks down at  $r=r_0$, and so the above
 solution is only local.  A global solution was proposed by gluing
 this one to the analogous solution on an anti-brane.\cite{Callan:1997kz},\cite{Gib}  The
 global solution
 then represents a wormhole connecting  the brane with an anti-brane a   distance of
 $2|X_{p+1}(r_0)|$  away, with a
 throat of minimum radius $r_0$ . 
The gluing of the two local solutions to form a wormhole occurs at
 $r=r_0$, precisely where there
 is a singularity
 in the electric field, which  might be a matter of concern.   
On the other hand, the  electric field singularity is a coordinate singularity, which is easily seen by
 transforming to another gauge.  Take for example the gauge where the
 $r-$coordinate of the static gauge is replaced by $z=X_{p+1}$.  In
 the new gauge the
 solution 
 is described by the inverse, call it  $R(z)$, of the function
 $X_{p+1}(r)$.  Now   the electrostatic field is in the $z-$direction,
 and there  are
 coordinate singularities at the location of the brane (and
 anti-brane).
We denote the electric field  ($\times\; 2\pi\alpha'$) in the new gauge
 by $E(z)$.  It can be computed locally by 
 performing a coordinate transformation from the static gauge,
\be E(z) = \frac{\partial R}{\partial z}\; f(R(z)) = \frac 1{g(R(z))}\; f(R(z)) \ee
Substituting in  the  solution for $f$ and $g$ given in
(\ref{sln0}) gives a constant electric field
\be E(z) = Q/C\; \label{eqoc}\ee
So in this 
coordinate frame there are no
singularities in the electric field (for $C\ne 0$).  
In  appendix A  we write down the
 action  in this gauge and show that (\ref{eqoc}) solves the
 corresponding equations of motion.

>From (\ref{xro}) it follows that there is a minimum separation distance
 between the brane and anti-brane for a fixed $Q$ (and $p>2$).  It is  equal to \be \min  \;  2| X_{p+1}(r_0)| = \frac
 {2 \sqrt{\pi(p-1)}} { (p-2)^{\frac{3p-4}{2p-2}}}\;\; \frac{
   \Gamma(\frac{2p-1}{2p-2} )}{
\Gamma(\frac{p}{2p-2})}\; |Q|^{\frac 1{p-1}}
   \;, \label{mindis}\ee and occurs when $C$ and $Q$ are
 constrained by
\be  C^2 = (p-1) Q^2\ee
 Below we plot the separation distance versus
 throat size for a fixed $Q$ when $p=3$:

\bigskip
\input epsf
\def\epsfsize#1#2{0.8#1}
\centerline{\hss{\epsffile{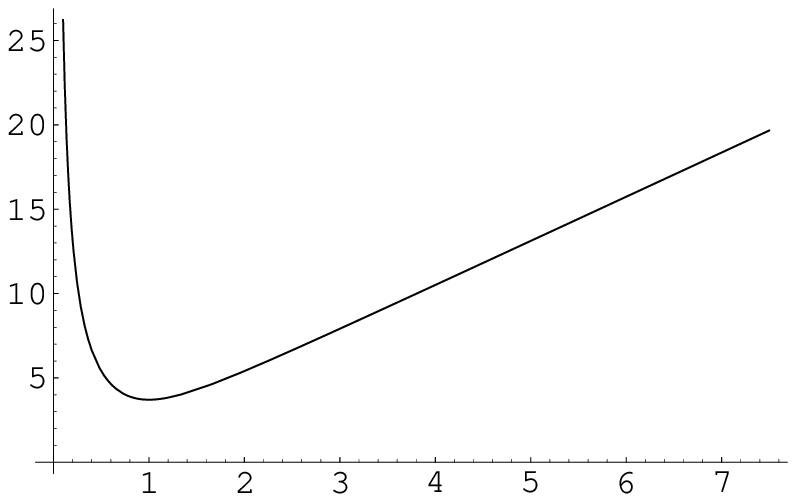}}}

\medskip
\centerline{\qquad\qquad\qquad
${\tt fig. 4}\qquad p=3\qquad Q=1\qquad  2|X_{p+1}(r_0)|\quad{\rm
vs.}\quad r_0 $}
\noindent

\subsection{Self-Energy}

Concerning the energy, one can apply the canonical formalism starting
  from the Lagrangian in (\ref{Dpact}).   For this again assume the
  static gauge and hence (\ref{hisg}).  The Hamiltonian density is 
\be {\cal H}^{(0)}_{DBI} =P^\alpha \dot X_{\alpha} + \Pi^\mu \dot {\cal A}_\mu - {\cal
  L}^{(0)}_{DBI}\;,\ee  where the dot denotes a
  time derivative and
\beqa P^\alpha &=& -\frac12  \sqrt{-\det h}\; (h^{0\mu}+h^{\mu 0}) \partial_\mu X ^\alpha
  \cr & &\cr
 \Pi^\mu &=& -\pi\alpha' \sqrt{-\det h}\; (h^{\mu 0}-h^{0\mu})
  \;,\label{mca}\eeqa   are the  momenta conjugate to $ X_\alpha$ and  ${\cal
  A}_\mu$, respectively.  As usual, the momentum
 conjugate to ${\cal
  A}_0$ is constrained to be zero.  After integrating by parts
\be {\cal H}^{(0)}_{DBI} =
P^\alpha \dot X_{\alpha} + \Pi^i F_{0i} +\sqrt{-\det h}-1 -\partial_i
  \Pi^i {\cal A}_0 \;, \label{ham}\ee where $i=1,2,...,p$.   The
  coefficient of $ {\cal A}_0$ gives the Gauss law constraint.  The
  remaining  terms are equal to
\be - \sqrt{-\det h}\; h^{00} \; -\;1\;\label{slfngydn} \ee
Although the Lagrangian density is invariant under the $SO(1,1)$
  symmetry (\ref{so11}), the Hamiltonian density is not.  Then unlike  the
  integral of the Lagrangian density, the  integral of the energy density will not be
  constant along the orbits of $SO(1,1)$.

 The
  integral of (\ref{slfngydn}) gives the self-energy of the
 DBI solutions (\ref{sln0}).
    After removing  a hole around the origin of
  radius $r$ in the integration domain one gets 
\be   {\cal U}(r) = \frac
 {T_p \Omega_{p-1}} {g_s}\; {\cal E}(r)\;,\qquad  {\cal E}(r) =\int^\infty_r dr  
 \frac{( r^{2p-2}   +{Q^2})} {\sqrt{ r^{2p-2} +Q^2-C^2}}-\int^\infty_0 drr^{p-1}
  \;,\label{efbih} \ee  where  the factor  $T_p/g_s$ comes from (\ref{Dpact}).
In the second integral we subtract off the total vacuum energy  of
  the brane.   Note that we must restrict the lower limit  in the
  first integral to be greater than or equal to $ r_0$ for the case
  $|C|>|Q|$.  The result can be expressed in terms of $X_{p+1}(r)$:
\be {\cal E}(r)\;  = \;-[(p-1)Q^2 +C^2]\; \frac {X_{p+1}(r)}{ p\;C} -
  \frac rp
\sqrt{r^{2p-2} +Q^2-C^2} \label{Eitor}
\ee  This gives a positive answer for the self-energy since ${X_{p+1}(r)}/{C}$ is negative for the  solutions, while the second term vanishes
  after evaluating at the minimum value of $r$ ($0$ for $|Q|\ge |C|$
  and $r_0$ for  $|C|>|Q|$).

For case   $i)\;|Q|> |C|$, one gets the total self-energy of the
  solution by setting $r$ in (\ref{Eitor}) to zero, which   yields
\be {\cal U}(0)\;  =\; \frac
 {T_p \Omega_{p-1}} {g_s}\; \frac{  \Gamma(\frac{3p-4}{2p-2}
 )\Gamma(\frac 1{2p-2}) }      {p (p-2)\sqrt{\pi}} \;
\frac{  (p-1)Q^2 +C^2}   { (Q^2 -C^2)^{\frac{p-2}{2p-2} } }\;,   \qquad{\rm for}\;p>2 
 \label{spen}\ee For a fixed $Q$ it goes monotonically from the
  Born-Infeld value  \be  {\cal U}_{\rm Born-Infeld}\;  = \;  \frac
 {T_p \Omega_{p-1}} {g_s}\; \frac{p-1  }      {p (p-2)\sqrt{\pi}} \;   \Gamma\biggl(\frac{3p-4}{2p-2}
 \biggr)\Gamma\biggl(\frac 1{2p-2}\biggr) \; { |Q|^{\frac{p}{p-1} } }\;, \qquad{\rm for}\;p>2
 \label{biengy}\ee corresponding to $C=0$, to
 infinity in the BPS limit, corresponding to $|C|\rightarrow |Q|$.
 We plot below ${\cal E}$  for  $Q=1$ and  $p=3$:

\bigskip
\input epsf
\def\epsfsize#1#2{0.8#1}
\centerline{\hss{\epsffile{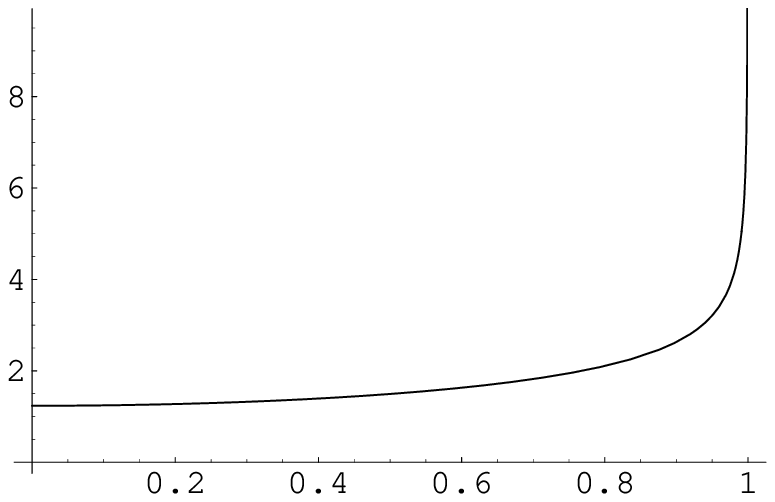}}}

\medskip
\centerline{\qquad\qquad\qquad
${\tt fig. 5}\qquad p=3\qquad Q=1\qquad {\cal E}\quad{\rm
vs.}\quad C\quad(C<Q) $}
\noindent 

For case $ii)\;|Q|=|C|$ the total self-energy is infinite.  At large distances  $|X_{p+1}|$,
the energy per unit length  of the infinite string solution is
constant.  From (\ref{Eitor}) it follows that  \be \frac{d {\cal
    U}}{d|X_{p+1}|}\rightarrow   \frac
 {T_p \Omega_{p-1}} {g_s}\; |Q|\;,\qquad{\rm  as} \quad|X_{p+1}| \rightarrow \infty\ee

For   case   $iii)\;|Q|< |C|$,  one gets the total self-energy  by setting $r$ in (\ref{Eitor}) equal to $r_0$:
\be {\cal U}(r_0)\;  = \frac
 {T_p \Omega_{p-1}} {g_s}\;\frac
 {\sqrt{\pi}} {p-2}\; \frac{
   \Gamma(\frac{2p-1}{2p-2} )}{
\Gamma(\frac{p}{2p-2})} \; \biggl(\frac{Q^2}{r_0^{p-2}} + \frac {r_0^p}p     \biggr)\;, \qquad{\rm for}\;p>2 \;\label{enxro}\ee 
For a fixed $Q$, the minimum energy configuration occurs for  $r_0^{2p-2} = (p-2)
  Q^2$, corresponding to the minimum separation distance (\ref{mindis})
  between the brane and anti-brane.
The minimum value for  $ {\cal U}(r_0)$ is \be {\cal U}_{\rm min}\;  =  \frac
 {T_p \Omega_{p-1}} {g_s}\;\; \frac
 {2\sqrt{\pi}(p-1)} {p(p-2)^{\frac{3p-4}{2p-2}}}\; \frac{
   \Gamma(\frac{2p-1}{2p-2} )}{
\Gamma(\frac{p}{2p-2})} \; |Q|^{\frac{p}{p-1}}\;, \qquad{\rm
for}\;p>2\label{wrhlen}\ee   If $Q=0$ the minimum  energy configuration
occurs when  the brane and anti-brane coincide.  For $Q\ne 0$ and a
   separation distance greater than the minimum value
  (\ref{mindis}), there are two possible  solutions with different
  throat sizes.  The one
  with a smaller throat is energetically favored.    Call $ {\cal U}_0(X_{p+1})$ and $ {\cal U}_1(X_{p+1})$ the energy of the
  thin and fat wormholes, respectively.   For a large separation distance,
 \beqa {\cal U }_0(X_{p+1}) &\rightarrow & \frac
 {T_p \Omega_{p-1}} {g_s}\;\;|QX_{p+1}| \cr & &\cr
 {\cal U}_1(X_{p+1})&\rightarrow & \frac
 {T_p \Omega_{p-1}} {pg_s}\;\;\biggr\{\frac
 {\sqrt{\pi}} {p-2}\; \frac{
   \Gamma(\frac{2p-1}{2p-2} )}{
\Gamma(\frac{p}{2p-2})}\biggl\}^{1-p} \;|X_{p+1}|^p \;
\eeqa   Upon plotting the
  energy versus the separation distance one gets a double-valued
  function, with a cusp at the minimum
  separation, as is illustrated below for  $Q=1$ and  $p=3$:
 
\bigskip
\input epsf
\def\epsfsize#1#2{0.8#1}
\centerline{\hss{\epsffile{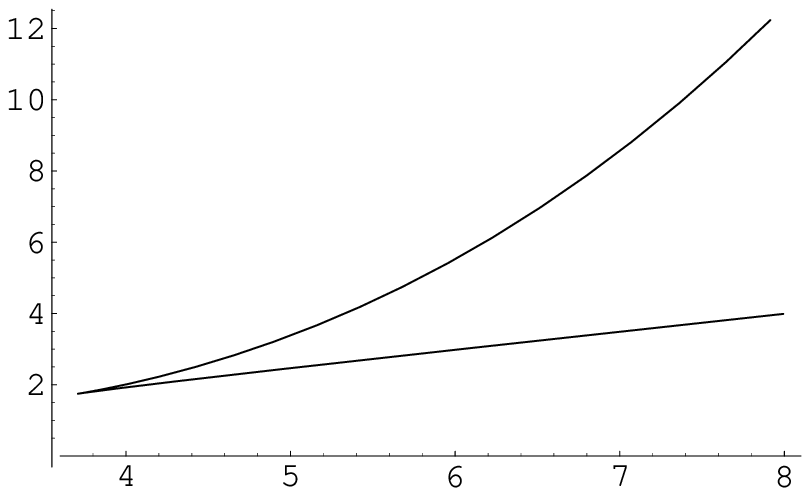}}}

\medskip
\centerline{\qquad\qquad\qquad
${\tt fig. 6}\qquad p=3\qquad Q=1\qquad {\cal E}\quad{\rm
vs.}\quad 2 |X_{p+1}(r_0)| $}

The minimum energy solution for fixed $Q$
  in  case $i) \;|Q|> |C|$ was the original Born-Infeld solution, while
  in case  $iii)\;|Q|< |C|$ it corresponded to (\ref{wrhlen}).  In both cases
  the energy goes like $ |Q|^{\frac{p}{p-1}}$.
  Assuming charge
  conservation, such solutions are energetically unstable under fission into far separated solutions with total
  charge equal to $Q$.   It was however pointed out in \cite{Gib} that
  fission may not be realized at the classical level for singular field configurations,
  and the above solutions are of this type.  Assuming fission does
  occur, either classically or quantum mechanically the minimum energy configuration should
   be an ensemble of  far separated wormholes  in  case   $iii)$ or
  Born-Infeld  solutions in case  $i)$ with the fundamental
  charge.   In comparing   $i)$
  with      $iii)$,  the ratio $ {\cal U}_{\rm min}/ {\cal U}_{\rm Born-Infeld}$ is
  less than one for $p\ge 4$, while it is greater than one for $p=3$.
  Thus for  $p\ge 4$, it is energetically more favorable for  wormholes
  to develop between  a charged brane and equally charged
  anti-brane than for Born-Infeld  configurations to develop on the brane
  and anti-brane.  The opposite is true for $p=3$.

\subsection{Thermodynamic considerations}

Here we make a side remark concerning the thermodynamics of wormholes.
Once again, for   case   $iii)$ when the energy is greater
 than the minimum, two types of wormholes with different thickness may be present.  Say they
 are in a heat bath with temperature $T$ and call $ \rho_0$ and
$ \rho_1$ the density of the
  thin and fat wormholes, respectively. If one assumes they are
 in dissipative and thermal equilibrium, then the ratio
of their densities at a temperature $T$ is given by \be \frac{\rho_1}{\rho_0} = \exp
\frac{{\cal U}_0(X_{p+1})  -{\cal U}_1(X_{p+1})}{k_B T}\ee

\section{Inclusion of  First Order Corrections}
\setcounter{equation}{0}

Here we  examine what happens to the zeroth order classical solutions upon
including the  first order  derivative corrections in the action.   We
already checked in \cite{Karatheodoris:2002bb} that
  the zeroth order Born-Infeld solution ($C=0$) does not survive upon
  the inclusion of such corrections.  More specifically, we numerically
  found 
 a classical solution to the corrected field equations, but 
 it was associated with an infinite value
 for the Lagrangian.  Because as with zeroth order, the Lagrangian is  $SO(1,1)$
 invariant, the result that the of an infinite value for the Lagrangian  follows for the entire orbit of solutions
 connected to the $C=0$ solution; i.e. case $i)$.  On the other hand,
 the case $ii)$ BPS solution ($|Q|=|C|$) is stable upon inclusion of
 the first order corrections, and just like at zeroth order,
the Lagrangian vanishes.  In fact the BPS solution is  known to survive   to all
  orders in the derivative expansion.\cite{Thorlacius:1997zd}  We shall
verify that this result is consistent with the explicit expression for
the first order terms obtained in \cite{Wyl},\cite{das}.   The
stability analysis for the wormhole solutions case $iii)$
 leads to the same results we obtain for case $i)$.
 Namely,  corrections to the zeroth order  solution lead to an
 infinitely large correction of the Lagrangian.

 The first
order corrections were initially  computed in \cite{Wyl},\cite{das} for the space-filling
D$9-$brane. A dimensional reduction  could  then
be performed to get the corrections to the DBI  action (\ref{Dpact})
for an arbitrary D$p-$brane.   We first  briefly recall the results of
the dimensional reduction procedure   at
zeroth order.\cite{Johnson}   One starts with the Born-Infeld (BI) action ${\cal S}^{(0)}_{BI}$ for  the space-filling
D$9-$brane.  It is written in terms of a $10\times 10$ matrix $\tilde h$ with elements
\be \tilde h_{AB} =\eta_{AB} +2\pi\alpha'\; F_{AB}\;,\qquad A,B,...=0,1,...,9\label{hAB}\ee where $
F_{AB}=\partial_A{\cal A}_B-\partial_B{\cal A}_A$ is the ten
dimensional field
strength  and we again assume   a flat background metric
$\eta_{AB}$. $\tilde h$ in 
(\ref{hAB}) can be obtained from $h$ in (\ref{giffh}) by assuming the
static gauge, which here means
$X^A=\xi^A$, for all $A$.   ${\cal S}^{(0)}_{BI}$
 is given by\be {\cal S}^{(0)}_{BI} = \frac{T_9}{ g_s} \int d^{10}\xi\; {\cal L}^{(0)}_{BI} \;,\qquad {\cal L}^{(0)}_{BI} = 1-
\sqrt{-\det [\tilde h_{AB}]} \label{biac}\;, \ee
and from  (\ref{Tp}),  $T_9 =1/{(4\pi^2 \alpha ')^5 }$. 
 In dimensional reduction
to the D$p-$brane,  $9-p$ of the nine spatial directions are
`T-dualized'.   Choose  the T-dual directions  to be  $A=
\alpha= p+1,p+2,...,9$. One of the  consequence of this procedure, is that the
 gauge potentials   ${\cal A}_\alpha $  in the T-dual directions get
 replaced  with the   transverse modes  $X_\alpha$ of the  D$p-$brane
according to \be 2\pi\alpha'{\cal A}_\alpha \rightarrow X_\alpha\;.\ee 
  The fundamental degrees of
freedom are then 
$X_\alpha$ and the remaining $p+1$ gauge potentials ${\cal A}_\mu$,
$\mu=0,1,...,p$, which are functions of the $p+1$
coordinates  of the brane  $\xi^\mu$.  Then the nonvanishing matrix elements
of $\tilde h$ are  $ \tilde h_{\mu\nu}$ and they are identical to $
h_{\mu\nu}$ of the DBI action written in the static gauge and
given in (\ref{hisg}).
Finally after performing integrations in the  T-dual directions (\ref{biac}) gets replaced by \be {\cal S}^{(0)}_{BI} =  \frac{T_p}{ g_s} \int d^{p+1}\xi\;\biggl( 1-\sqrt{-\det [\tilde h_{\mu\nu}]}\biggr) \ee
where the D$p-$brane
tension is again given by (\ref{Tp}), and one  recovers (\ref{Dpact}) in the
static gauge.
So instead of working with $  h_{\mu\nu}$ as we did in the previous
section we could have started with  the $10\times 10$ matrix $ \tilde h$.
Then for  the static spherically symmetric solutions of the previous
section where just the $p+1$
transverse mode is excited,  $$  2\pi \alpha'F_{0i}=f  \hat r_i\qquad  2\pi \alpha'F_{p+1\;i}=g  \hat r_i \;,\qquad
i=1,2,...p\;,$$ where $\hat r$ is the unit vector in the radial direction and
spherical symmetry means  $ f $ and  $g  $ are only functions  of the
radial variable $r$.  The   $10\times 10$ matrix  
 $ \tilde h$ takes the form
\be \tilde h= \pmatrix{-1  & - f \hat r  & & \cr
 f \hat r  &\BI_{p\times p} &  g \hat r & \cr
 &- g  \hat r & 1 & \cr & &  & &\BI_{(8-p)\times(8-
   p)}\cr}\;\label{hinv1}\ee

The first
order corrections ${\cal S}^{(1)}_{BI}$ to the action $ {\cal
  S}^{(0)}_{BI}$ of the space-filling D$9-$brane obtained in \cite{Wyl},\cite{das}
 involve first
and second derivatives of the field strength
$F_{AB}$.  They  are contained in the rank-$4$ tensor
\be  S_{ABCD} =2\pi \alpha'\partial_A \partial_B
F_{CD}
+(2\pi \alpha')^2 \tilde h^{EG}( \partial_A F_{CE} \partial_B
F_{DG} -   \partial_A F_{D E} \partial_B
F_{CG} ) \;, \ee which is antisymmetric in the last two indices.
Here $\tilde h^{AB}\tilde h_{BC}= \delta^A_C$.  
The total action is
$${\cal  S}^{(0)}_{BI} + {\cal S}^{(1)}_{BI} = \frac{T_9}{ g_s} \int
  d^{10}x\;\biggl\{  1-
\sqrt{-\det [\tilde h_{AB}]} \;\biggl(1 +\frac \kappa 4  \Delta\biggl)\biggr\} \;, $$
\be \Delta =\; \tilde h^{AB} \tilde h^{CD}\tilde h^{EG}  \tilde h^{IJ}(
S_{B CEG} S_{D AIJ} - 2  S_{GI BC}
S_{J E D A} )\;,\label{Dlta} \ee where  $\kappa ={{
    (2\pi\alpha')^2}\over {48}} $.
We again    specialize to the case
where a single transverse mode [the  $(p+1)^{\rm th}$ mode] is excited on
a $p\le 8$ brane, and consider static spherically symmetric fields.   So the ansatz  for  $\tilde h$ is again (\ref{hinv1}).
Its determinant and inverse are given 
by 
\be \det\tilde h= -1 +  f^2 -  g^2  \label{deth}\ee   and
\be \tilde h^{-1}=\frac1{\det\tilde  h} \pmatrix{1+ g^2 & \hat r f  &
  - f g &\cr -\hat r f  & -\BI + (f^2-g^2) P & \hat r g & \cr
 - f g 
 &- \hat r g &  f^2-1 & \cr  & &  &\det\tilde  h \;\BI_{(8-p)\times(8- p)} \cr
}\;,\ee respectively.   $P$ is the projection matrix $P_{ij} = \delta_{ij}
-\hat r_i \hat r_j$, satisfying $P_{ij} \hat r_j=0$ and $P_{ij}P_{jk}
=P_{ik}$.  Some work shows that the nonvanishing components of $ S_{ABCD}$ are
\beqa
 S_{ijk0}=- S_{ij0k}&=& \det \tilde h\; {H_f}'\; \hat r_i\hat r_j \hat
 r_k -\frac1r \biggl(H_f+\frac fr\biggr)(P_{ik}\hat r_j
+P_{jk}\hat r_i) + \biggl(\frac fr \biggr)' P_{ij}\hat r_k \cr & &\cr
S_{ijk\;p+1}=- S_{ij\;p+1\;k}&=&\det \tilde h\;{ H_g}'\; \hat r_i\hat r_j \hat
 r_k -\frac1r \biggl(H_g +\frac gr \biggr)(P_{ik}\hat r_j
+P_{jk}\hat r_i) + \biggl(\frac gr \biggr)' P_{ij}\hat r_k \cr & &\cr
S_{ijk\ell}&=& \frac{(\ln \det \tilde  h)'}{2r}(P_{ik} \hat r_j \hat r_\ell
 +
P_{j\ell} \hat r_i \hat r_k -P_{i\ell} \hat r_j \hat r_k-P_{jk} \hat r_i \hat r_\ell)\cr & &\cr & &
+\;\frac{1 +(\det\tilde  h)^{-1}}{r^2} (P_{ik} P_{j\ell} -
P_{i\ell}P_{jk})
\;,
\eeqa
where \be H_f = \frac{f'}{\det\tilde  h}\;,\qquad   H_g =
\frac{g'}{\det\tilde  h}
\ee
the prime here denoting derivatives in $r$.  In addition we define
$$ H_k= f H_g - g H_f $$
Substituting into the formula in  (\ref{Dlta}) for $\Delta$ gives
\beqa \Xi = - \frac14 (\det \tilde h)^2 \Delta &=&  {H_f}'^2 - {H_g}'^2 +
{H_k}'^2 \cr& &\cr
& + &\frac{p-1}{r^2}\biggl\{\biggl(2+(\det\tilde  h)^2 \biggr)\;({H_f}^2 - {H_g}^2 + {H_k}^2) \cr & & \cr
& +&\frac1{2} (\ln\det\tilde  h)'^2  +\frac2{r} (\ln\det\tilde  h)'  -\frac1{r}
(\det\tilde  h)' \cr& &\cr & 
+&
\frac1{r^2}(1+\det\tilde  h)\biggl( p+1+(p-2)\det\tilde h\biggr)\biggr\}\;\eeqa
So for the above ansatz the correction to the zeroth order  
Lagrangian density ${\cal L}^{(0)}_{DBI}$ in (\ref{Dpact}) is
\be {\cal L}^{(1)}_{DBI} =  \frac{ \kappa\;\Xi}{(-\det\tilde h)^{3/2}}  \;
\label{radL}\ee

 In obtaining the equations of motion one
must again write $f(r)$ and $g(r)$ in terms of potentials and extremize with
respect to  the latter.  As the general system is quite involved, below  we shall restrict to  functions
  $f(r)$ and $g(r)$ which are related by a constant factor
\be  \frac{f(r)}Q  = \frac{g(r)}C\;, \ee as what occurred for the
zeroth order solutions (\ref{sln0}).     This
set of configurations
respects the  $SO(1,1)$ 
  symmetry (\ref{so11}) and (\ref{so11qc}).   Using
\be   \frac{H_f(r)}Q =\frac{H_g(r)}C\;,\qquad H_k(r)  = 0 \ee the
Lagrangian density simplifies, and it is    $SO(1,1)$ invariant.
 Once again there are  three distinct orbits: $i)\;|Q|>|C|$, $ii)\;|Q|=|C|$ and $iii)\;|Q|<|C|$, and one expects that these orbits are classified by the corresponding  value for the spatial
  integral of the Lagrangian density, which is now $ {\cal
  L}^{(0)}_{DBI} + {\cal
  L}^{(1)}_{DBI}$.   To compute the latter we only have to examine one
point on each of  the orbits, which we do below.  

 $i)\;|Q|>|C|$.  A convenient point on this orbit is the purely
 electrostatic case, where the transverse mode is suppressed:
 $f(r)=2\pi\alpha' {\cal A}_0'(r)$ and    $g(r)=0$.  Substituting this into $ {\cal
  L}^{(0)}_{DBI} + {\cal
  L}^{(1)}_{DBI}$,
 and varying with respect to ${\cal A}_0(r)$ gives the corrected Born-Infeld equation
$$ \frac{ \sqrt{-\det \tilde
  h}} \kappa \biggl( r^{p-1}f-Q\sqrt{-\det \tilde
  h}  \biggr)$$
\be =   \biggl[ \frac{2r^{p-1} H_f' } {(-\det\tilde
  h)^{3/2}}\biggr]''  - \frac{ 3r^{p-1} f H_f'^2 } {(-\det\tilde h)^{3/2}}
-\frac{2(p-1)(3+f^4)(r^{p-3} H_f)'}{(-\det\tilde h)^{3/2}}\label{feco}\ee
$$+
\frac{(p-1)r^{p-5}f} {(-\det\tilde h)^{3/2}}\biggl\{ f^4(p-2 -r^2 H_f^2)+
f^2(-2p-3+4r^2 H_f^2)+ 3(-2p +6 +3r^2 H_f^2)  \biggr\}$$
 To get back the zeroth order equations set the left hand side equal
 to zero.  So the right hand side represents  the
derivative corrections.  The zeroth order Born-Infeld solution
satisfies (\ref{feco}) as $r\rightarrow\infty$, so the corrections are
negligible  in this region.    In \cite{Karatheodoris:2002bb},
starting with  the zeroth order Born-Infeld
solution at $r\rightarrow\infty$, we used
(\ref{feco}) to numerically integrate  to finite $r$.  We found the
resulting corrections to $f(r)$ at finite $r$ to be small, and just
like at zeroth order,  $f$ tends to $1$ as $r\rightarrow 0$.   Call
$f_0(r)$  the zeroth solution, and  $f_1(r)$ the correction it receives at first order.   Below we
plot $f_0(r)$ and  $f_0(r)+f_1(r)$ when
$Q=1,\;\; C=0$ and $p=3$:

\bigskip
\input epsf
\def\epsfsize#1#2{0.8#1}
\centerline{\hss{\epsffile{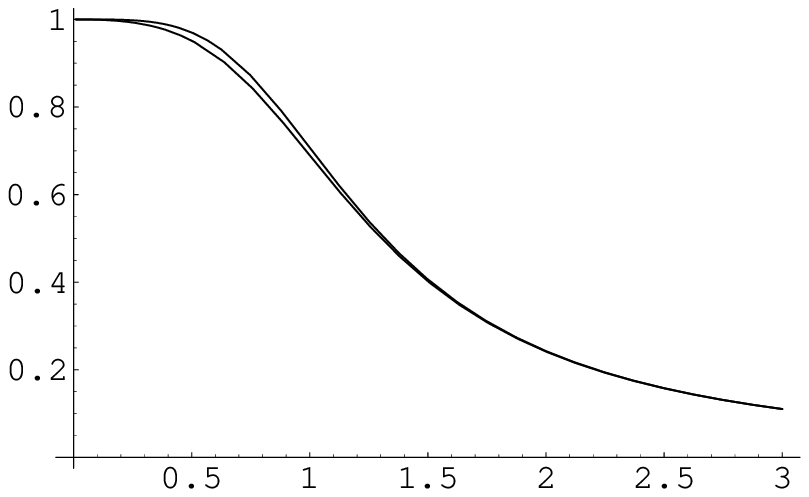}}}

\medskip
\centerline{\qquad\qquad\qquad
${\tt fig. 7}\qquad  f_0\;\; {\tt  and }\;\; f_0+f_1\;\;{\tt
  vs}\;\;r$}
\centerline{\qquad\qquad\qquad
$\qquad\qquad{\tt for}\;\;  p=3,\;\; Q=1,\;\;C=0 $}

\noindent
In the above we set $\kappa=1$ which is equivalent to choosing the
scale for $r$.    If the zeroth order Born-Infeld solution gives a reasonable
approximation to a classical solution in the full effective theory,
 and one can apply the derivative expansion to get the latter, then higher order corrections should be small.   In particular, we expect only a
small change in the value of the Lagrangian at the next order.   If
one assumes
 this to be the case  
a Taylor expansion about the zeroth order solution gives
$$ \int d^{p}\xi\;[ {\cal
  L}^{(0)}_{DBI} + {\cal
  L}^{(1)}_{DBI}](f_0+f_1)  $$ \be \approx   \int d^{p}\xi\;[ {\cal
  L}^{(0)}_{DBI}](f_0 )  +   \int d^{p}\xi\;[{\cal
  L}^{(1)}_{DBI}](f_0 ) +\int d^{p}\xi\;\frac{\delta  {\cal
  L}^{(0)}_{DBI}}{\delta f}\bigg|_{f=f_0 }\; f_1 \ee
The last term vanishes by the field equations, and so the first order
correction to the Lagrangian is $ \int d^{p}\xi\;[{\cal
  L}^{(1)}_{DBI}](f_0 ) $.  However, it is easy to check that  $[{\cal
  L}^{(1)}_{DBI}](f_0 )$ diverges near the origin as $1/r^{3p+1}$.  So
 the first order correction  is not small;
 Rather,  $ \int d^{p}\xi\;[{\cal
  L}^{(1)}_{DBI}](f_0 ) $ is infinite!  This agrees with the
result found in   \cite{Karatheodoris:2002bb},  and indicates that the Born-Infeld solution, and
indeed all case $i)$ solutions, are unstable under inclusion of first
order derivative  corrections.

  $ii)\;|Q|=|C|$.  This is the   BPS case  $g(r)=f(r)$.   Here   $ {\cal L}^{(1)}_{DBI}=0$, and so just like at zeroth
    order  the Lagrangian
vanishes.   To find equations of motion we must first vary $f$ and $g$
(or more precisely   $X_{p+1}$
and ${\cal A}_0$) separately and then impose the
BPS condition.  We do this in Appendix B.  (Actually, there we don't
    impose the restriction of spherical symmetry.)  The result is simply
\be \nabla^2\; \biggl\{ 1 + 2\kappa \;(\nabla^2)^2 \biggr\}\;{\cal
    A}_0 =0 \;,\label{eqfbps}\ee with the  same equation for  $X_{p+1}$.
For the case of spherically symmetric solutions we can use
$\nabla^2= \frac1{r^{p-1}} \partial_r r^{p-1}\partial_r$.
Eq. (\ref{eqfbps}) says that the zeroth order solution is also valid
at  first order.  This agrees with   \cite{Thorlacius:1997zd}, where it was shown that the
BPS solution is valid to all orders.  The result (\ref{eqfbps}) thus
    provides a  check of the computations   in \cite{Wyl},\cite{das}.

$iii)\;|Q|<|C|$.  A convenient point is the purely transverse case.  Here the electric field vanishes:
 $f(r)=0$ and    $g(r)=X_{p+1}'(r)$.  Substituting this in (\ref{radL})
 and varying $X_{p+1}(r)$ gives
$$ \frac{ \sqrt{-\det\tilde
  h}} \kappa \biggl( r^{p-1}g-C\sqrt{-\det\tilde
  h}  \biggr)$$
\be =   \biggl[ \frac{2r^{p-1} H_g' } {(-\det\tilde
  h)^{3/2}}\biggr]''  + \frac{ 3r^{p-1} g H_g'^2 } {(-\det\tilde h)^{3/2}}
-\frac{2(p-1)(3+g^4)(r^{p-3} H_g)'}{(-\det\tilde h )^{3/2}}\label{fectw}\ee
$$+
\frac{(p-1)r^{p-5}g} {(-\det\tilde  h)^{3/2}}\biggl\{ g^4(p-2 +r^2 H_g^2)+
g^2(2p+3+4r^2 H_g^2)+ 3(-2p +6 -3r^2 H_g^2)  \biggr\}$$
 Again   to get back the zeroth order equations set the left hand side equal
 to zero, and so the right hand side represents  the
derivative corrections.  (\ref{fectw})
 is also obtained by making the  transformation
$f(r) \rightarrow ig(r)$ and $Q \rightarrow iC$  in (\ref{feco}), so
solving for real $g(r)$   in (\ref{fectw}) is equivalent to solving
for imaginary $f(r)$  in (\ref{feco}).  As with case $i)$, starting with  the zeroth order
solution at $r\rightarrow\infty$, we can  use
(\ref{fectw}) to numerically integrate  to finite $r$.  We  find that just
like at zeroth order,  $g$ becomes singular at some finite $r$, which
appears to be slightly greater than $r_0$.  We can then conclude that
the corrections cause the wormhole to become wider.   Call
$g_0(r)$ the zeroth order solution,  and  $g_1(r)$ the correction it
receives at first order.   Below we
plot $g_0(r)$ and $g_0(r)+ g_1(r)$ when 
$C=1,\; Q=0$ and $p=3$:

\bigskip
\input epsf
\def\epsfsize#1#2{0.8#1}
\centerline{\hss{\epsffile{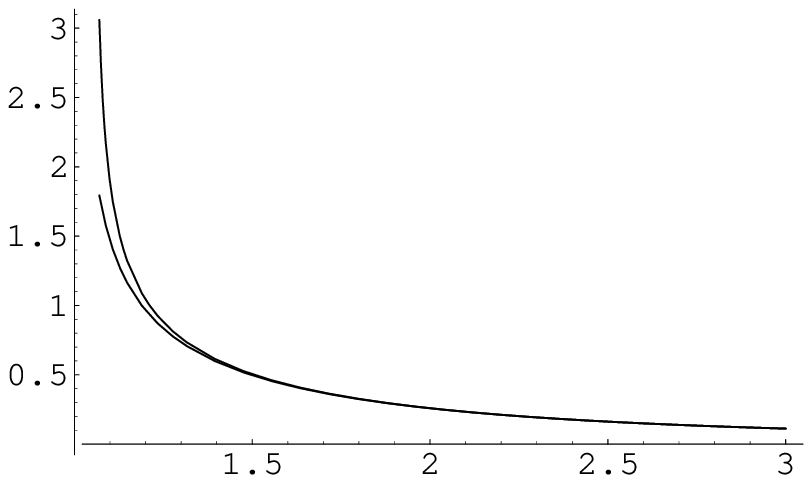}}}

\centerline{\qquad\qquad\qquad
${\tt fig. 8} \qquad  g_0\;\; {\tt  and }\;\;g_0+ g_1\;\;{\tt
  vs}\;\;r $}
\centerline{\qquad\qquad\qquad
$\qquad\qquad{\tt for}\;\;  p=3,\;\; C=1,\;\;Q=0 $}

\noindent
Again we set $\kappa=1$.  Figure  8 shows that the correction $g_1$ to the
solution is small away
from the wormhole throat.   On the other hand, the corresponding
correction to the Lagrangian density appears not to be small, as is
indicated below where we numerically compare $[{\cal
  L}^{(0)}_{DBI}+ {\cal
  L}^{(1)}_{DBI}](g_0+g_1)  $ and $[ {\cal
  L}^{(0)}_{DBI}](g_0)  $: 

\bigskip
\input epsf
\def\epsfsize#1#2{0.8#1}
\centerline{\hss{\epsffile{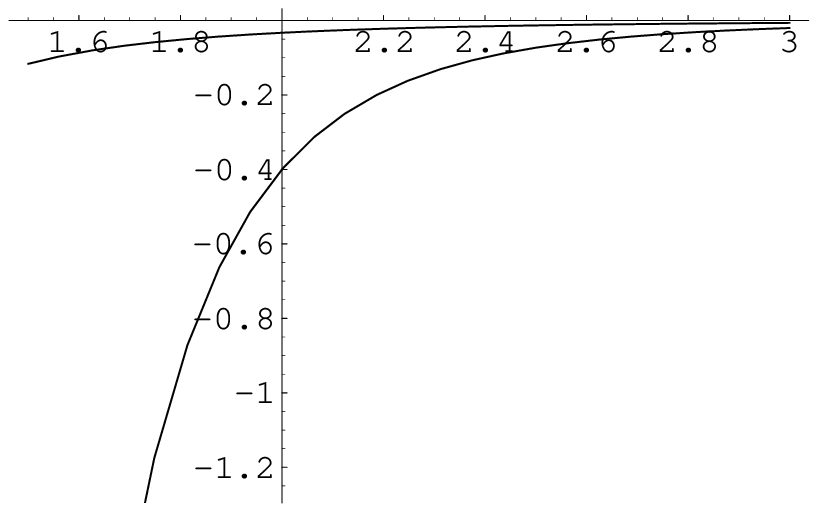}}}

\medskip
\centerline{\qquad\qquad\qquad
${\tt fig. 9} \qquad  [ {\cal
  L}^{(0)}_{DBI}](g_0) \; {\tt  and }\; [ {\cal
  L}^{(0)}_{DBI}+ {\cal
  L}^{(1)}_{DBI}](g_0+g_1) \;\;{\tt  vs}\;\;r $}
\centerline{\qquad\qquad\qquad
$\qquad\qquad{\tt for}\;\;
p=3,\;\; C=1,\;\;Q=0 $}

\noindent
 The lower curve in figure 9 is $ [ {\cal
  L}^{(0)}_{DBI}+ {\cal
  L}^{(1)}_{DBI}](g_0+g_1)  $.  We then expect large corrections to its
  integral.  In fact, we find that the numerical
  integration of   $ [ {\cal
  L}^{(0)}_{DBI}+ {\cal
  L}^{(1)}_{DBI}](g_0+g_1) $ fails to give a convergent result.  So just
  as in case $i)$, the integral of the Lagrangian density appears to
  be ill-defined,  indicating that  the case $iii)$ solutions
  are also unstable under inclusion of first
order derivative  corrections.  

\section{Conclusion}

 The preliminary indications here are that the classical wormhole
 solution may not be a reasonable approximation to a solution in the
 full $D-$brane theory.  If so it therefore cannot prevent the
 decay of the brane-anti-brane system.  More generally, it appears
 that the  only
 solution that survives higher order derivative corrections may be the
 BPS solution.  On the other hand, a more extensive analysis may be possible.  For example, it would be interesting to drop the
 restriction to static solutions.  Perhaps time dependent
 configurations can survive first order derivative
 corrections, or perhaps the zeroth order static solutions evolve to time
 dependent ones after including the higher order.   To check this would require
 combining two separate stability analyses, which we referred to as $a)$ and $b)$
 in the introduction.  A final but
 unpleasant (from a computational point of view)  possibility is that 
solutions are recovered only  after going beyond the first order.
 Moreover, all orders may be required, meaning the   solutions may be non perturbative.

\bigskip

{\parindent 0cm{\bf Acknowledgement}}                                         
 
This work was supported in part by the joint NSF-CONACyT grant
E120.0462/2000.

\bigskip
\noindent
\appendice $\qquad$ {\bf Alternative Gauge}

\setcounter{equation}{0}

Here we  reconstruct the zeroth order wormhole solutions  starting with the
 action written in an alternative gauge.  This gauge is obtained by replacing the
 $r-$coordinate of the static gauge  by $z=X_{p+1}$ gauge.  It has the
 advantage that it removes the coordinate singularity appearing at the midway
point  of the  wormhole, and shows that there is
 a smooth solution connecting the brane to anti-brane.  Coordinate
 singularities re-appear, however, at the location of the brane and anti-brane. 
We show that the electric field is well behaved in this gauge, and in
 fact is  a constant.

   Consider the
domain $S^{p-1}\times{\mathbb{R}}^2$, with  local coordinate patches
used to parametrize the
D$p-$brane.  Denote by $z$ one of the coordinates of ${\mathbb{R}}^2$, while the other
corresponds to time $t$.  We look for spherically symmetric static solutions with  
\be X_0 =t\;,\qquad X_1^2 + X_2^2 +\cdot\cdot\cdot+X_p^2 =
R(z)^2\;,\qquad X_{p+1} =z \ee  
So for example for the 
D$3-$brane we can write \beqa X_1&=& R(z) \sin\theta \cos\phi \cr
X_2&=& R(z) \sin\theta \sin\phi \cr
X_3&=& R(z) \cos\theta  \;, \eeqa
using standard spherical coordinates $\theta$ and $\phi$,
\be h_{tt} =-1\;,\qquad h_{\theta\theta}= R(z)^2\;,\qquad
h_{\phi\phi}= R(z)^2\sin^2\theta \;,\qquad h_{zz}= R'(z)^2+1\;,\ee
where here the prime denotes a derivative in $z$. 
Now introduce a $z-$dependent electrostatic field in the $z-$direction,
leading to the off-diagonal components
\be h_{tz}=-h_{zt} =- E(z) \ee
In terms of the electrostatic potential ${\cal
  A}_0$ which we now write as a function of $z$,   $E(z)=2\pi\alpha'\partial_z{\cal
  A}_0(z)$.
 After performing
the angular integrations,  the DBI Lagrangian  $L_{DBI}^{(0)}$ (ignoring the vacuum term) will be proportional to $R(z)^2 \;\sqrt{R'(z)^2-E(z)^2+1}$.
Generalizing to  arbitrary $p$,  
\be L_{DBI}^{(0)}\; \propto \;  R(z)^{p-1} \;\sqrt{R'(z)^2-E(z)^2+1}\ee
Variations in  the electrostatic potential and $R(z)$  give 
\be \biggl(\frac{ R(z)^{p-1} \;E(z)}{\sqrt{R'(z)^2-E(z)^2+1}}
\biggr)^{'} =0 \ee \be R''(z) R(z) = (p-1)\; (R'(z)^2-E(z)^2 +1)\;, \ee respectively.  From the first equation 
\be\bigg| \frac {E(z)}Q\bigg| = \sqrt{\frac{R'(z)^2 +1}{R(z)^{2p-2} +Q^2}}\;,\label{ezdir}\ee
where $Q$ is an integration constant, and substituting into the second equation
\be R''(z) = (p-1)\;R(z)^{2p-3}\; \frac{R'(z)^2 +1}{R(z)^{2p-2} +Q^2} \ee
After integrating once
\be R'(z) =\frac 1C \sqrt{ R(z)^{2p-2} - C^2+Q^2}\;,\label{rprm}\ee where
$C$ is an integration constant.  Since $R'(z)$ corresponds to
$1/g(r)$, the result agrees with
(\ref{sln0}).  For the wormhole solution, $R(z)$ is nonsingular everywhere 
except at the location  of the brane and anti-brane.   At the
midway-point  on the wormhole, $R$ is a well-defined  function of $z$,
and  is a minimum since
\be R''(z_{mid}) = \frac{p-1}{r_0}\;\biggl(1-\frac {Q^2}{C^2}\biggr)\;,
\qquad  R(z_{mid})=r_0=(C^2-Q^2)^{\frac 1{2p-2}}\;, \ee
and $|Q|<|C|$  for wormhole solutions.
 So now
coordinate singularities appear at the brane and anti-brane, rather
than at the midway-point on the wormhole.   By substituting
(\ref{rprm}) into  (\ref{ezdir}) one gets that the electric field
$E(z)$ in the $z-$direction is a constant
\be | E(z)| = \bigg| \frac QC\bigg|\;, \ee  which agrees with
(\ref{eqoc}).  It goes to one in the
BPS limit, and $1/\sqrt{p-1}$ for the minimum energy wormhole.  We
conclude that  in this 
coordinate frame there are no
singularities in the electric field (for $C\ne 0$).

\bigskip
\noindent
\appendice   $\qquad$  {\bf  First order BPS equations}

\setcounter{equation}{0}

Here we derive the first order BPS equation (\ref{eqfbps}).  Unlike in
section 3, we make
no restriction to spherical symmetry.  Our starting point is then not
(\ref{hinv1}), but
\be \tilde h= \pmatrix{-1  & - \vec f   & & \cr
 \vec f  &\BI_{p\times p} &  \vec g  & \cr
 &- \vec g  & 1 & \cr & &  & &\BI_{(8-p)\times(8-
   p)}\cr}\;\label{ghinv1}\;,\ee 
where $\vec f=2\pi \alpha'\vec \nabla {\cal A}_0$ and $\vec g=\vec \nabla X_{p+1}$ are vector fields on the D$p-$brane.
The general  BPS condition is $\vec f=\vec g$.  Upon imposing this 
condition on $\tilde h^{-1}$ one gets 
 \be \tilde h^{-1}|_{BPS}= \pmatrix{-1-\vec f^2 &-\vec f  &
  \vec f^2 & &\cr \vec f  & \BI_{p\times p} & -\vec f & \cr
 \vec f^2  & \vec f & -\vec f^2+1 & \cr& &  & &\BI_{(8-p)\times(8-
   p)}\cr 
}\;,\ee while the only nonvanishing components of $ S_{ABCD}$ are 
\be   S_{ijk0}|_{BPS}= S_{ijk\;p+1}|_{BPS}=- S_{ij0k}|_{BPS}=- S_{ij\;p+1\;k}|_{BPS}=
\partial_i\partial_j f_k \ee  Since the BPS action vanishes,  we must impose 
 $\vec f =\vec g$ {\it after} performing the variations in the action to find the BPS field equations, i.e. we
 must be allowed to perform separate variations of  ${\cal A}_0$ and $ X_{p+1}$.
By varying  ${\cal A}_0$ and then setting  $\vec f=\vec g$ , \beqa
 \delta
\tilde h^{00}|_{BPS}&=& -2\;(1+\vec f^2) \;  \vec f\cdot\delta\vec f
 \cr \delta\tilde  h^{p+1\;p+1}|_{BPS}&= &-2\;\vec f^2\; \vec f\cdot\delta\vec f
  \cr  \delta
\tilde h^{0\;p+1}|_{BPS}&=& (1+2\vec f^2) \;  \vec f\cdot\delta\vec f
\; =\;\delta \tilde h^{p+1\;0}|_{BPS}  \cr 
\delta \tilde h^{i0}|_{BPS}&=& (1+\vec f^2)\;\delta f_i+ \vec f
\cdot\delta\vec f\;f_i\;=\;-\delta\tilde  h^{0i}|_{BPS}\cr
\delta\tilde 
h^{i\;p+1}|_{BPS}&= & -\vec f^2\;\delta f_i- \vec f
\cdot\delta\vec f\;f_i\; =\;-\delta
\tilde h^{p+1\;i}|_{BPS} \cr \delta
\tilde h^{ij}|_{BPS}&=& - f_i\delta
 f_j + f_j\delta f_i \;, \eeqa while the nonvanishing components of $
 \delta S_{ABCD}$ are \beqa   \delta S_{ijk0}|_{BPS}&=&-\delta S_{ij0k}|_{BPS}=
\partial_i\partial_j\delta f_k
+ \delta S_{ijk\;p+1}|_{BPS} \cr & &\cr
\delta S_{ijk\;p+1}|_{BPS}&=&-\delta S_{ij\;p+1\;k}|_{BPS}=
\partial_i f_\ell \;\partial_j f_k\;\delta f_\ell
- \tilde h^{0\ell} \partial_i \delta f_k\; \partial_j f_\ell
+(i\rightleftharpoons j)\cr & &\cr \delta S_{ijk\ell}|_{BPS}&=&\partial_j f_\ell\;
\partial_i \delta f_k+
\partial_i f_k\;\partial_j \delta f_\ell -(i\rightleftharpoons j)
 \eeqa
In evaluating $\delta \Delta|_{BPS}$ we can use $(\tilde h^{AB} S_{ijAB})|_{BPS}=0$.
Then
\beqa \delta \Delta|_{BPS}&=& -4\tilde h^{CD} S_{jiDA}\; \{ \delta
\tilde h^{AB} S_{ijBC}
+ \tilde h^{AB} \delta S_{ijBC} +\tilde  h^{AB} \delta \tilde h^{i\ell}S_{\ell jBC}
\}\;|_{BPS}\cr & &\cr
&=& -8\;
\partial_i\partial_j f_k\;\partial_i\partial_j\delta f_k \eeqa
Now substitute $\vec f=2\pi \alpha'\vec \nabla {\cal A}_0$ to  obtain
(\ref{eqfbps})  from the variation of ${\cal A}_0$ in  ${\cal  L}^{(0)}_{DBI} + {\cal
  L}^{(1)}_{DBI}$.  One gets the same results from variations of $X_{p+1}$.

\end{document}